\documentclass[twocolumn,showpacs,preprintnumbers,amsmath,amssymb,prl]{revtex4}
\usepackage{dcolumn}
\usepackage{bm}
\usepackage[dvips]{graphicx}
\usepackage{wrapfig}
\usepackage{psfrag}
\begin{document}

\def\sh{\mathop{\rm sh}\nolimits}
\def\ch{\mathop{\rm ch}\nolimits}
\def\var{\mathop{\rm var}}\def\exp{\mathop{\rm exp}\nolimits}
\def\Re{\mathop{\rm Re}\nolimits}
\def\Sp{\mathop{\rm Sp}\nolimits}
\def\kp{\mathop{\text{\ae}}\nolimits}
\def\bk{{\bf {k}}}
\def\bp{{\bf {p}}}
\def\bq{{\bf {q}}}
\def\lra{\mathop{\longrightarrow}}
\def\Const{\mathop{\rm Const}\nolimits}
\def\sh{\mathop{\rm sh}\nolimits}
\def\ch{\mathop{\rm ch}\nolimits}
\def\var{\mathop{\rm var}}
\def\mK{\mathop{{\mathfrak {K}}}\nolimits}
\def\mR{\mathop{{\mathfrak {R}}}\nolimits}
\def\mv{\mathop{{\mathfrak {v}}}\nolimits}
\def\mV{\mathop{{\mathfrak {V}}}\nolimits}
\def\mD{\mathop{{\mathfrak {D}}}\nolimits}
\def\mN{\mathop{{\mathfrak {N}}}\nolimits}
\def\mS{\mathop{{\mathfrak {S}}}\nolimits}

\newcommand\ve[1]{{\mathbf{#1}}}

\def\Re{\mbox {Re}}
\newcommand{\Z}{\mathbb{Z}}
\newcommand{\R}{\mathbb{R}}
\def\mK{\mathop{{\mathfrak {K}}}\nolimits}
\def\mR{\mathop{{\mathfrak {R}}}\nolimits}
\def\mv{\mathop{{\mathfrak {v}}}\nolimits}
\def\mV{\mathop{{\mathfrak {V}}}\nolimits}
\def\mD{\mathop{{\mathfrak {D}}}\nolimits}
\def\mN{\mathop{{\mathfrak {N}}}\nolimits}
\def\ml{\mathop{{\mathfrak {l}}}\nolimits}
\def\mf{\mathop{{\mathfrak {f}}}\nolimits}
\newcommand{\ccm}{{\cal M}}
\newcommand{\cE}{{\cal E}}
\newcommand{\cV}{{\cal V}}
\newcommand{\cI}{{\cal I}}
\newcommand{\cR}{{\cal R}}
\newcommand{\cK}{{\cal K}}
\newcommand{\cH}{{\cal H}}

\def\br{\mathop{{\bf {r}}}\nolimits}
\def\bS{\mathop{{\bf {S}}}\nolimits}
\def\bA{\mathop{{\bf {A}}}\nolimits}
\def\bJ{\mathop{{\bf {J}}}\nolimits}
\def\bn{\mathop{{\bf {n}}}\nolimits}
\def\bg{\mathop{{\bf {g}}}\nolimits}
\def\bv{\mathop{{\bf {v}}}\nolimits}
\def\be{\mathop{{\bf {e}}}\nolimits}
\def\bp{\mathop{{\bf {p}}}\nolimits}
\def\bz{\mathop{{\bf {z}}}\nolimits}
\def\bbf{\mathop{{\bf {f}}}\nolimits}
\def\bb{\mathop{{\bf {b}}}\nolimits}
\def\ba{\mathop{{\bf {a}}}\nolimits}
\def\bx{\mathop{{\bf {x}}}\nolimits}
\def\by{\mathop{{\bf {y}}}\nolimits}
\def\br{\mathop{{\bf {r}}}\nolimits}
\def\bs{\mathop{{\bf {s}}}\nolimits}
\def\bH{\mathop{{\bf {H}}}\nolimits}
\def\bk{\mathop{{\bf {k}}}\nolimits}
\def\be{\mathop{{\bf {e}}}\nolimits}
\def\bnul{\mathop{{\bf {0}}}\nolimits}
\def\bq{{\bf {q}}}

\newcommand{\oV}{\overline{V}}
\newcommand{\vkp}{\varkappa}
\newcommand{\os}{\overline{s}}
\newcommand{\opsi}{\overline{\psi}}
\newcommand{\ov}{\overline{v}}
\newcommand{\oW}{\overline{W}}
\newcommand{\oPhi}{\overline{\Phi}}

\def\mI{\mathop{{\mathfrak {I}}}\nolimits}
\def\mA{\mathop{{\mathfrak {A}}}\nolimits}

\def\st{\mathop{\rm st}\nolimits}
\def\tr{\mathop{\rm tr}\nolimits}
\def\sign{\mathop{\rm sign}\nolimits}
\def\d{{\mathrm d}}
\def\const{\mathop{\rm const}\nolimits}
\def\O{\mathop{\rm O}\nolimits}
\def\Spin{\mathop{\rm Spin}\nolimits}
\def\exp{\mathop{\rm exp}\nolimits}
\def\mU{\mathop{{\mathfrak {U}}}\nolimits}
\newcommand{\cU}{{\cal U}}
\newcommand{\cD}{{\cal D}}

\def\mI{\mathop{{\mathfrak {I}}}\nolimits}
\def\mA{\mathop{{\mathfrak {A}}}\nolimits}

\def\st{\mathop{\rm st}\nolimits}
\def\tr{\mathop{\rm tr}\nolimits}
\def\sign{\mathop{\rm sign}\nolimits}
\def\const{\mathop{\rm const}\nolimits}
\def\O{\mathop{\rm O}\nolimits}
\def\Spin{\mathop{\rm Spin}\nolimits}
\def\exp{\mathop{\rm exp}\nolimits}

\title{Neutrino oscillations: another physics?}

\author {S.N. Vergeles\vspace*{4mm}\footnote{{e-mail:vergeles@itp.ac.ru}}}

\affiliation{Landau Institute for Theoretical Physics,
Russian Academy of Sciences,
Chernogolovka, Moskow region, 142432 Russia \linebreak
and   \linebreak
Moscow Institute of Physics and Technology, Department
of Theoretical Physics, Dolgoprudnyj, Moskow region,
Russia}

\begin{abstract} It is shown that the neutrino oscillations phenomenon
may be attributed to the Wilson fermion doubling phenomenon. The
Wilson fermion doubling exists only on the lattices, both periodic
and non-periodic (simplicial complexes). Just the last case plays a
key role here. Thereby, the neutrino oscillations may show for the
existence of a space-time granularity.
\end{abstract}


\maketitle

{\bf 1.} The neutrino oscillations, i.e. the mutual oscillating
transitions of the neutrinos of different generations, are observed
for a long time now. The common explanation of the phenomenon is
based on the assumption the neutrino mass matrix is non-diagonal.
Moreover, in order to match all the experimental evidences, it is
necessary to introduce the extra neutrino fields, which are sterile
regarding to all interactions (naturally, except gravitational one).
The sterile neutrinos cannot be observed directly: they are coupled
to the three known neutrino generations only by means of a common
mass matrix, and this is the way they give a contribution to the
neutrino oscillations. The introduction of sterile neutrinos does
not exhaust all difficulties of the theory. The detailed description
of the neutrino oscillations experiments and theory can by found,
for example, in \cite{1}, \cite{2}, and in numerous references
there.

It is shown here that another physics may provide the neutrino
oscillations.  The basis for this physics is the Wilson lattice
fermion doubling phenomenon \cite{3,31,32,33,34,35}. Here the Wilson doubling on
irregular lattices (simplicial complexes) is interesting (see
below). This case was studied in \cite{4}, \cite{5}. Thus first of
all I must outline shortly the physics of irregular (doubled)
fermion quanta on the "breathing" simplicial complexes \cite{5} in
the frame of discrete gravity (see also \cite{6}).

{\bf 2.} Let's describe shortly the model of discrete quantum
gravity which is determined on the simplicial complexes. For
simplicity, only the gravitational and fermionic dynamic variables
are introduced in this model. Since the mutual geometrical locations
of the vertices of the complex are determined by the gravitational
variables, the mutual geometrical locations of the vertices are
described by a wave function in the quantum theory. Therefore the
simplicial complex can be named as "breathing" one here.

Note that the considered model of discrete quantum gravity is not
realistic for some reasons. Nevertheless, maybe certain features of
the model adequately simulate a Nature character. In this work I
make the assumption that the fermion fields are determined on a
"breathing" 4-dimensional simplicial complex. This assumption is
basic here.

Further all definitions and designations are similar to that in
\cite{5,6}. The quantities related with Euclidean signature are supplied by the
additional lower index $(E)$. For example, the Euclidean Dirac
matrices $(4\times4)$
\begin{gather}
\gamma_{(E)}^a\gamma_{(E)}^b+\gamma_{(E)}^b\gamma_{(E)}^a=2\delta^{ab},
\quad
\gamma_{(E)}^5=\gamma_{(E)}^1\gamma_{(E)}^2\gamma_{(E)}^3\gamma_{(E)}^4\,,
\quad
\nonumber \\
\tr\,\gamma_{(E)}^5\gamma_{(E)}^a\gamma_{(E)}^b\gamma_{(E)}^c\gamma_{(E)}^d=4\,
\varepsilon_{(E)}^{abcd}. \label{dqg80}
\end{gather}
For each oriented 1-simplex $a_ia_j$ of the 4-dimensional simplicial
complex $\mK$ an element of the group $\Spin(4)$
\begin{gather}
\Omega_{ij}=\Omega^{-1}_{ji}=\exp\left(\frac{1}{2}\omega^{ab}_{(E)ij}
\sigma_{(E)}^{ab}\right), \quad
\sigma_{(E)}^{ab}=\frac14\left[\gamma_{(E)}^a,\gamma_{(E)}^b\right]
\label{dqg90}
\end{gather}
and an element of the Clifford algebra
\begin{gather}
\hat{e}_{(E)ij}\equiv
e^a_{(E)ij}\gamma_{(E)}^a\equiv-\Omega_{ij}\hat{e}_{(E)ji}\Omega_{ij}^{-1},
\label{dqg100}
\end{gather}
are assigned. The elements (\ref{dqg90}) belong to the compact group
if the  variables $\omega^{ab}_{(E)ij}$ are real. Let the index $A$
enumerates 4-simplices, $a_{Ai},\,a_{Aj},\,a_{Ak},\,a_{Al}$, and
$a_{Am}$ be all five vertices of a 4-simplex with index $A$ and
$\varepsilon_{Aijklm}=\pm 1$ depending on whether the order of
vertices $a_{Ai}a_{Aj}a_{Ak}a_{Al}a_{Am}$ defines the positive or
negative orientation of this 4-simplex. Later the notations
$a_{Ai},\,a_{Aj},\,\Omega_{Aij}$ and so on indicate that 1-simplex
$a_ia_j$ belong to 4-simplex with index $A$. The action of the
gravitational and Dirac fields associated with four-dimensional
simplicial complex $\mK$ has the form
\begin{gather}
\mA=\frac{1}{5\times
24}\sum_A\sum_{i,j,k,l,m}\varepsilon_{Aijklm}\tr\,\gamma_{(E)}^5
\times
\nonumber \\
\times\left\{\frac{2}{l^2_P}\Omega_{Ami}\Omega_{Aij}\Omega_{Ajm}
\hat{e}_{(E)Amk}\hat{e}_{(E)Aml}-\right.
\nonumber \\
\left.-\frac{1}{24}\hat{\Theta}_{Ami}
\hat{e}_{(E)Amj}\hat{e}_{(E)Amk}\hat{e}_{(E)Aml}\right\}\,,
\label{dqg110}
\end{gather}
\begin{gather}
\hat{\Theta}_{Aij}=
\frac12\gamma_{(E)}^a\left(\psi_{Ai}^{\dag}\gamma_{(E)}^a
\Omega_{Aij}\psi_{Aj}-\psi_{Aj}^{\dag}\Omega_{Aji}\gamma_{(E)}^a\psi_{Ai}\right).
\label{dqg120}
\end{gather}

The partition function for a discrete gravity is defined as follows:
\begin{gather}
\mU\sim\bigg(\prod_{\cE}\int\d\Omega_{\cE}\int\d
e_{\cE}\bigg)\left(\prod_{\cV}\int\d\psi^{\dag}_{\cV}
\d\psi_{\cV}\right)\exp(i\mA). \label{dqg140}
\end{gather}
The indexes $\cV$ and $\cE$ enumerate all vertexes and 1-simplexes,
correspondingly, $\d\Omega_{\cE}$ is the Haar measure on the group
$\Spin(4)$, $\d e_{\cE}=\prod_a\d e_{\cE}^a$.

Let's observe the formal transition to the continuum physics with
Minkowski signature. To pass to the Minkowski signature one must
deform the integration paths in (\ref{dqg140}) and make the
substitutions as follows ($\alpha,\,\beta,\,\ldots=1,\,2,\,3$)
\begin{gather}
\omega_{(E)\,ij}^{4\alpha}=i\,\omega_{ij}^{0\alpha}, \quad
\omega_{(E)\,ij}^{\alpha\beta}= -\omega_{ij}^{\alpha\beta},
\nonumber \\
e_{(E)\,ij}^4=i\,e_{ij}^{0}, \quad
e_{(E)\,ij}^{\alpha}=-e_{ij}^{\alpha},
\nonumber \\
\gamma_{(E)}^4=\gamma^0, \quad \gamma_{(E)}^{\alpha}=i\,
\gamma^{\alpha}, \quad \psi_{(E)i}^{\dag}=\opsi=\psi^{\dag}\gamma^0.
\label{dqg250}
\end{gather}
The variables $\omega_{ij}^{ab}$ and $e_{ij}^a$ in the right hand
sides of Eqs. (\ref{dqg250}) are real. So we have
\begin{gather}
\omega^{ab}_{(E)ij} \sigma_{(E)}^{ab}=\omega^{ab}_{ij}\sigma_{ab},
\quad  \sigma^{ab}\equiv\frac14\left[\gamma^a,\gamma^b\right],
\label{dqg255}
\end{gather}
and the elements (\ref{dqg90}) of the compact group $\Spin(4)$
transform to the elements of the noncompact group $\Spin(3,1)$.
We introduce the local coordinates of the vertices \footnote{The
construction of the local coordinates can be realized as follows.
Let's begin with an immersion of any 4-simplex to a four dimensional
Euclidean space. Thus its vertexes acquire the Cartesian
coordinates. Then the same is to be done with all adjacent
4-simlices, so that the common vertexes remain common in the
Euclidean space, and so on. This process is possible since complex
$\mK$ is orientable.}:
$x^{\mu}_{A\,i}\equiv x^{\mu}(a_{A\,i}),$ where
$\mu=1,2,3,4$ and $\d x^{\mu}_{ji}\equiv x^{\mu}_{A\,i}-x^{\mu}_{A\,j}$.
Suppose we have a smooth fields $\omega^{ab}_{\mu}(x)$ and
$e^a_{\mu}(x)$ in the Euclidean space. Then the lattice variables
$\omega^{ab}_{ji}=\omega^{ab}_{\mu}(x)\d x^{\mu}_{ji}$,
$e^a_{mi}=e^a_{\mu}(x)\d x^{\mu}_{mi}$,
$ x=x_{A\,j}$
are determined by the 1-forms $\omega^{ab}_{\mu}(x)\d x^{\mu}_{ji}$
and $e^a_{\mu}(x)\d x^{\mu}_{ji}$. The inverse procedure is
considered in \cite{5,6}. The fermion field is a 0-form:
$\psi_i=\psi(x), \ x^{\mu}\equiv x^{\mu}_{A\,i}$. Thus, neglecting
high derivatives of the fields $\omega^{ab}_{(E)\,\mu}(x)$,
$e^a_{(E)\mu}(x)$ and $\psi(x)$, we pass on from the lattice action
(\ref{dqg110}) to the usual continual Minkowski gravity action in
the Palatini form up to the multiplier $(-i)$. This unnecessary
multiplier is easily removed by the the substitution $x^4=i\,x^0$,
where the coordinate $x^0$ is the real Minkowski time. Indeed, the
action in Palatini form is the integral of 4-form which is linear
and uniform relative to the $\d x^4$. Thus in the naive long-wave
limit and Minkowski signature we have the usual gravitational action in Palatini form.

{\bf 3.} Now I discuss shortly the problem of Wilson fermion
doubling phenomenon on the irregular lattice \cite{4,5}. For that
end the situation must be simplified as much as possible. Thus
further we believe that
\begin{gather}
\Omega_{ij}=1\,\,, \ \ \
\bigl(e_{ij}^a+e_{jk}^a+\ldots+e_{li}^a\bigr)=0. \label{cor350}
\end{gather}
Here the sum in the parentheses is taken on any closed path. Eqs.
(\ref{cor350}) mean that the curvature and torsion are equal to
zero, so that the geometrical realization of the $n$ Dimensional
complex $\mK$ is in the $n$ Dimensional Minkowski space, the
cartesian coordinates of a vertex $a_i$ have the values $x^a_i$ and
$e^a_{ij}=x^a_j-x^a_i$, $a=0,\ldots,n-1$. Here we are interested in
the cases $n=4,\,3$.

  \begin{figure}[t]
\psfrag{nuie}{\rotatebox{0}{\kern0pt\lower0pt\hbox{{$\nu
_e^{\cal{I}}$}}}}
\psfrag{nue}{\rotatebox{0}{\kern0pt\lower0pt\hbox{{$\nu_{e}$}}}}
\psfrag{nuimu}{\rotatebox{0}{\kern0pt\lower0pt\hbox{{$\nu
_{\mu}^{\cal{I}}$}}}}
\psfrag{numu}{\rotatebox{0}{\kern0pt\lower0pt\hbox{{$\nu_{\mu}$}}}}
\psfrag{bp}{\rotatebox{0}{\kern0pt\lower0pt\hbox{{$\bar{\psi}$}}}}
\psfrag{rl}{\rotatebox{0}{\kern0pt\lower0pt\hbox{{$\circlearrowright$}}}}
\psfrag{rp}{\rotatebox{0}{\kern0pt\lower0pt\hbox{{$\circlearrowleft$}}}}
\psfrag{lwf}{\rotatebox{0}{\kern0pt\lower0pt\hbox{{long-wavelength
fermions}}}}
\psfrag{2}{\rotatebox{0}{\kern0pt\lower0pt\hbox{{$2$}}}}
\psfrag{3}{\rotatebox{0}{\kern0pt\lower0pt\hbox{{$3$}}}}
\psfrag{4}{\rotatebox{0}{\kern0pt\lower0pt\hbox{{$4$}}}}
\psfrag{1}{\rotatebox{0}{\kern0pt\lower0pt\hbox{{$1$}}}}
\psfrag{irf}{\rotatebox{0}{\kern0pt\lower0pt\hbox{{irregular
fermions}}}}
\psfrag{gr}{\rotatebox{0}{\kern0pt\lower0pt\hbox{{graviton}}}}
\psfrag{5}{\rotatebox{0}{\kern0pt\lower0pt\hbox{{$5$}}}}
\psfrag{6}{\rotatebox{0}{\kern0pt\lower0pt\hbox{{$6$}}}}
    \center{\includegraphics[width=0.25\textwidth]{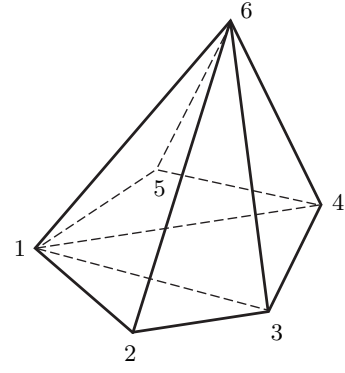}}
    \caption{An example of 2-dimensional complex with the fermion doubling property}
    \label{fig:1}
  \end{figure}

We need the lattice analog of the Dirac Hamiltonian.  For this let's
choose a spacelike $(n-1)D$ subcomplex $\mS$ of the $nD$ complex
$\mK$. It is assumed that the geometrical realization of the
subcomplex $\mS$ is in the hypersurface determined by the Eq.
$x^0=\const$. Thus, we have all $e^0_{ij}=0$ on $\mS$, and instead
of the $e^a_{ij}$ the quantities $e^{\alpha}_{pr}$ are used. The
vertices of subcomplex $\mS$ are designated as $a_p,\, a_q,\,
a_r,\,\ldots$. Equation for the zero eigenfunction of the Dirac
Hamiltonian looks like
\begin{gather}
-\frac{i}{(n-1)}\sum_{q(p)}\sum_{\alpha}S_{\alpha pq}
\gamma^0\gamma^{\alpha}\psi^{(0)}_{q}=0,
\label{cor430}
\end{gather}
\begin{gather}
S_{\alpha
pq}=\frac{1}{((n-2)!)^{2}}\sum_{A(p,q)}\sum_{\{r_1,\ldots,r_{n-2}\}}
\varepsilon_{\alpha\beta_1\ldots\beta_{n-2}}\times
\nonumber \\
\times\varepsilon_{A(p,q)pqr_1\ldots
r_{n-2}}e^{\beta_1}_{A(p,q)pr_1}\ldots
e^{\beta_{n-2}}_{A(p,q)pr_{n-2}}
\label{cor380}
\end{gather}
Here index $A(p,q)$ enumerates all $(n-1)$-simplexes containing a
fixed 1-simplex $a_pa_q$, index $q(p)$ enumerates all vertexes $a_q$
of the 1-simplexes $a_pa_q$ with a fixed vertex $a_p$,
$\varepsilon_{A(p,q)pqr_1\ldots r_{n-2}}=\pm1$  depending on the
orientation of the $(n-1)$-simplex $a_pa_qa_{r_1}\ldots
a_{r_{n-2}}$.

For simplicity let's consider the case $n=3$. Let $\mv_p$ denotes
the part of the subcomplex $\mS$ consisting of all 2-simplexes
having a common vertex $a_p$. We enumerate the vertexes
$a_{q(p)}\in\partial\mv_p$ so that the vertex $a_{q(p)+1}$ go after
the vertex $a_{q(p)}$ in tracing the boundary $\partial\mv_p$
anticlockwise, and we shall assume that the index $q(p)$ is
determined up to $(mod \,\,n)$, where $n$ is the number of vertexes
on $\partial\mv_p$. Introduce the complex coordinate
$z_p=x_p^1+ix_p^2$ for the vertex $a_p$. Let the spinor $\psi$ be
2-component, the upper component is designated as $\varphi$,
$\gamma^{\alpha}=\sigma^{\alpha},\,\alpha=1,2$. Eq. (\ref{cor430})
splits, and for upper component we have \cite{4}:
\begin{equation}
\sum_{q(p)}\,z_{q(p)}\,(\varphi_{q(p)+1}-\varphi_{q(p)-1})=0.
\label{cor385}
\end{equation}

The Wilson doubling means that the system of equations
(\ref{cor385}) has the irregular solutions $\varphi^{{\cal
I}}_{p,q}\equiv\varphi^{{\cal I}}_p-\varphi^{{\cal I}}_q\neq 0$,
$|\varphi^{{\cal I}}_{p,q}|\sim|\varphi^{{\cal I}}_p|$ for the next
vertexes $a_p$ and $a_q$.

In \cite{4} the 2-D complexes are constructed which do not possess
the Wilson doubling property. But such a complexes are not to be
seen as typical one, but most probably as exclusive complexes. For
example, the 2-D complex with 6 vertexes and $S^2$-topology (see
Fig.~\ref{fig:1}) possess the Wilson doubling property. Many other examples
also demonstrate this property. Unfortunately, I have not classified
the simplicial complexes with respect to the Wilson doubling
property. But being guided by the experience I shall assume that
almost all simplicial complexes manifest the Wilson doubling.

%
%

  \begin{figure}[t]
\psfrag{nuie}{\rotatebox{0}{\kern0pt\lower0pt\hbox{{$\nu
_e^{\cal{I}}$}}}}
\psfrag{nue}{\rotatebox{0}{\kern0pt\lower0pt\hbox{{$\nu_{e}$}}}}
\psfrag{nuimu}{\rotatebox{0}{\kern0pt\lower0pt\hbox{{$\nu
_{\mu}^{\cal{I}}$}}}}
\psfrag{numu}{\rotatebox{0}{\kern0pt\lower0pt\hbox{{$\nu_{\mu}$}}}}
\psfrag{x}{\rotatebox{0}{\kern0pt\lower0pt\hbox{{$x$}}}}
\psfrag{y}{\rotatebox{0}{\kern0pt\lower0pt\hbox{{$y$}}}}
\psfrag{z}{\rotatebox{0}{\kern0pt\lower0pt\hbox{{$z$}}}}
\psfrag{w}{\rotatebox{0}{\kern0pt\lower0pt\hbox{{$w$}}}}
\psfrag{xinf}{\rotatebox{0}{\kern0pt\lower0pt\hbox{{$x_{\infty}$}}}}
\psfrag{yinf}{\rotatebox{0}{\kern0pt\lower0pt\hbox{{$y_{\infty}$}}}}
\psfrag{bp}{\rotatebox{0}{\kern0pt\lower0pt\hbox{{$\bar{\psi}$}}}}
\psfrag{rl}{\rotatebox{0}{\kern0pt\lower0pt\hbox{{$\circlearrowright$}}}}
\psfrag{rp}{\rotatebox{0}{\kern0pt\lower0pt\hbox{{$\circlearrowleft$}}}}
\psfrag{lwf}{\rotatebox{0}{\kern0pt\lower0pt\hbox{{long-wavelength
fermions}}}}
\psfrag{2}{\rotatebox{0}{\kern0pt\lower0pt\hbox{{$2$}}}}
\psfrag{3}{\rotatebox{0}{\kern0pt\lower0pt\hbox{{$3$}}}}
\psfrag{4}{\rotatebox{0}{\kern0pt\lower0pt\hbox{{$4$}}}}
\psfrag{1}{\rotatebox{0}{\kern0pt\lower0pt\hbox{{$1$}}}}
\psfrag{irf}{\rotatebox{0}{\kern0pt\lower0pt\hbox{{irregular
fermions}}}}
\psfrag{gr}{\rotatebox{0}{\kern0pt\lower0pt\hbox{{graviton}}}}
\psfrag{B8}{\rotatebox{0}{\kern0pt\lower0pt\hbox{{$\alpha_{\ve{x}{+}\ve{e}_3}$}}}}
\psfrag{B9}{\rotatebox{0}{\kern0pt\lower0pt\hbox{{$\gamma_{\ve{x}{+}\ve{e}_1}$}}}}
    \center{\includegraphics[width=0.35\textwidth]{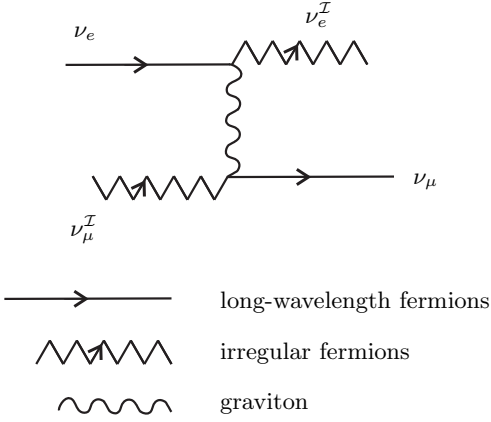}}
    \caption{The process of the electron neutrino transition to the muon one.}
    \label{fig:2}
  \end{figure}

{\bf 4.} Let $\{\psi^{{\cal I}(0)}_{s\,j}\}, s=1,\,2,\,\ldots,\,S$
be the complete independent set of the irregular zero modes. Each of
them satisfy Eq. (\ref{cor430}) on a $3D$ subcomplex $\mS$ and the
discrete Dirac equation with Minkowski signature on the complex
$\mK$. The soft irregular modes can be conceived of as $\psi^{{\cal
I}}_j\{g\}=\sum_sg^{{\cal I}}_{s\,j}\psi^{{\cal I}(0)}_{s\,j}$,
where the fields $g^{{\cal I}}_{s\,j}$ are slowly varying and
satisfying the equation
\begin{gather}
i\sum_{s'}\alpha^b_{i\,ss'}\partial_bg^{{\cal I}}_{s'\,i}=0,
\label{cor480}
\end{gather}
\begin{gather}
\alpha^b_{i\,ss'}=\frac{1}{n(n+1)}\left[\sum_a\sum_{j(i)}\left(\opsi^{{\cal
I}(0)}_{s\,i}S_{a\,ij}e^b_{ij}\gamma^a\psi^{{\cal
I}(0)}_{s'\,j}\right)\right].
\label{cor490}
\end{gather}
Here the quantity $S_{a\,ij}$ is constructed in the same way as in
(\ref{cor380}) but on the $4D$ complex $\mK$. Now the irregular part
of the Dirac field is expressed as
\begin{gather}
\psi^{{\cal I}}_i=\sum_sg^{{\cal I}}_{si}\psi^{{\cal I}(0)}_{s\,i},
\label{cor500}
\end{gather}
where the fields $g^{{\cal I}}_{si}$ are the Grassmann variables.
Here our main interest is the chronological correlator
\begin{eqnarray}\label{cor510}
&&iS_c^{{\cal I}}(x_i-x_j)\equiv\langle \hat{T}\psi^{{\cal
I}}_i\opsi\,{}^{{\cal I}}_j\rangle_{\psi,\,e}=
\\&&\hskip-20pt \nonumber
=\psi^{{\cal
I}(0)}_{s\,i}\big{\langle}\big(i\alpha^a\partial_a\big)^{-1}_{ss'\,ij}\big{\rangle}_e\,
\opsi^{{\cal I}(0)}_{s'\,j}\sim
a^2\gamma^a\partial_a\delta^{(4)}\left(x_i-x_j\right).
\end{eqnarray}
The subscripts $\psi,\,e$ mean the averaging over Grassmann  and
$e$-variables, correspondingly. Here $a$ acts as the effective
lattice sale. It is seen from (\ref{cor510}) that the irregular
quanta are "bad" quasiparticles. This is a consequence of the facts
that (i) the lattice is irregular and "breathing" and (ii) the
irregular quanta interact strongly with the lattice.
The result (\ref{cor510}) is important below, it has been obtained in \cite{5}.

{\bf 5.} Let us suppose that there are two left neutrino fields
$\nu_e$ and $\nu_{\mu}$ in the model \footnote{The model admits the
introduction of any number of the neutrino massless fields.}.
Suppose also that there are the nonzero densities $n^{{\cal
I}}_{\nu_e}$ and $n^{{\cal I}}_{\nu_{\mu}}$ of the irregular quanta
(calculated per volume of quanta number).

\begin{figure}[t]
\psfrag{nuie}{\rotatebox{0}{\kern0pt\lower0pt\hbox{{$\nu
_e^{\cal{I}}$}}}}
\psfrag{nue}{\rotatebox{0}{\kern0pt\lower0pt\hbox{{$\nu_{e}$}}}}
\psfrag{nuimu}{\rotatebox{0}{\kern0pt\lower0pt\hbox{{$\nu
_{\mu}^{\cal{I}}$}}}}
\psfrag{numu}{\rotatebox{0}{\kern0pt\lower0pt\hbox{{$\nu_{\mu}$}}}}
\psfrag{bp}{\rotatebox{0}{\kern0pt\lower0pt\hbox{{$\bar{\psi}$}}}}
\psfrag{rl}{\rotatebox{0}{\kern0pt\lower0pt\hbox{{$\circlearrowright$}}}}
\psfrag{rp}{\rotatebox{0}{\kern0pt\lower0pt\hbox{{$\circlearrowleft$}}}}
\psfrag{1}{\rotatebox{0}{\kern0pt\lower0pt\hbox{{1}}}}
\psfrag{2}{\rotatebox{0}{\kern0pt\lower0pt\hbox{{$2$}}}}
\psfrag{3}{\rotatebox{0}{\kern0pt\lower0pt\hbox{{$3$}}}}
\psfrag{4}{\rotatebox{0}{\kern0pt\lower0pt\hbox{{$4$}}}}
\psfrag{B5}{\rotatebox{0}{\kern0pt\lower0pt\hbox{{$\alpha_{\ve{x}}$}}}}
\psfrag{B6}{\rotatebox{30}{\kern0pt\lower0pt\hbox{{$\beta_{\ve{x}}$}}}}
\psfrag{B7}{\rotatebox{0}{\kern0pt\lower0pt\hbox{{$\gamma_{\ve{x}{+}\ve{e}_2}$}}}}
\psfrag{B8}{\rotatebox{0}{\kern0pt\lower0pt\hbox{{$\alpha_{\ve{x}{+}\ve{e}_3}$}}}}
\psfrag{B9}{\rotatebox{0}{\kern0pt\lower0pt\hbox{{$\gamma_{\ve{x}{+}\ve{e}_1}$}}}}
    \center{\includegraphics[width=0.32\textwidth]{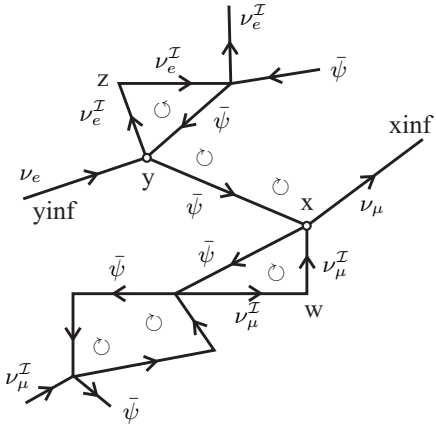}}
    \caption{The process of electron neutrino transition to the muon one \textquotedblleft under the microscope\textquotedblright\ on the lattice}
    \label{fig:3}
  \end{figure}

Let's consider the scattering process pictured in Fig.~\ref{fig:2}.
In Fig.~\ref{fig:3} the same process is considered "under the microscope" (on
the lattice). One must take into account that after the calculation
of fermion integral in (\ref{dqg140}) the fermion segments with
arrows will be assigned to 1-simplices with following properties:
(i) the fermion segments form continuous broken lines on the
complex, so that the arrows are unidirectional; (ii) two arrows of
each kind of fermions come into each vertex and two arrows come out;
(iii) each fermion broken line can be closed or unclosed. In the
first case we have a vacuum fluctuation, in the latter case the line
describes propagation of a real fermion.

It seems that the propagation of a real irregular fermion on the
lattice is similar in a sence to the dynamics of a Brownian
particle. But in our case, due to the chaotic interference, the
propagator (\ref{cor510}) decreases at $|x_i-x_j|\rightarrow\infty$
more quickly than the probability of Brownian random walks for the
distance $|x_i-x_j|$. This analogy once more justifies Eq.
(\ref{cor510}).

The symbols $\circlearrowleft$ and $\circlearrowright$ in Fig.~\ref{fig:3}
mean the contribution from the gravity part of the action
(\ref{dqg110}) under high-temperature expansion. For long-wavelength
physics the high-temperature expansion does not work. The lines
$(y_{\infty}\rightarrow y)$ and $(x\rightarrow x_{\infty})$ denote
the long-wavelength quanta of $\nu_e$ and $\nu_{\mu}$ popagation,
correspondingly.

%
%
%

Let's estimate the amplitude of the process on Figs.~\ref{fig:2},~\ref{fig:3}. Note that
the asymptotic quanta $\nu_e$ and $\nu_{\mu}$ give the factor
$\left\{\exp\left(ik_ey-ik_{\mu}x\right)/\sqrt{\omega_e\omega_{\mu}}\right\}$
if the long-wavelength quanta are normalized as one quantum per unit
volume. Here $k$ and $\omega$ are the 4-momentum and frequency of
quanta. The irregular fermion quanta do not the asymptotic states,
therefore they give the contribution $\left\{iS_{c\mu}^{{\cal
I}}(x-w)\otimes iS_{c\,e}^{{\cal I}}(y-z)\right\}$. (The external
spinor states are not interesting here.) Thus the estimation for the
amplitude is as follows
\begin{gather}
iM\sim\int\d^4x\int\d^4y
\frac{\exp\left(ik_ey-ik_{\mu}x\right)}{\sqrt{\omega_e\omega_{\mu}}}\times
\nonumber \\
\times\left\{iS_{c\mu}^{{\cal I}}(x-w)\otimes iS_{c\,e}^{{\cal
I}}(y-z)\right\}\left\{-iGD^{{\cal I}}(x-y)\right\}.
\label{am10}
\end{gather}
Here $G$ is the Newton gravitational constant, $D^{{\cal I}}(x-y)$
means the graviton propagator of irregular (not long-wavelength)
quanta. The spinor or tensor indexes are not interesting here.
Obviously, the estimation
$
D^{{\cal I}}(x-y)\sim a^2\delta^{(4)}(x-y)
\label{am20}
$
is valid (compare with (\ref{cor510})). I shall assume that $a\sim
l_P$ and $\delta^{(4)}(0)\sim l_P^{-4}$, where $l_P$ is the Planck
scale, $G=l^2_P$. Since $|w-x|\sim|y-z|\sim l_P$, the integrations
over $y$ and $x$ in (\ref{am10}) give
\begin{gather}
iM\sim\frac{1}{l^2_P\sqrt{\omega_e\omega_{\mu}}}\int\d^4xe^{i(k_e-k_{\mu})x}
\sim\frac{\Delta x^0\delta^{(3)}({\bf k}_e-{\bf
k}_{\mu})}{l_P^2\omega}.
\label{am30}
\end{gather}
It is assumed that the translational invariance of the system is
valid. Here $\Delta x^0$ in (\ref{am30}) is of the order of the
process duration, $\Delta x^0\sim l_P$. Thus, omitting the
$\delta$-function, we obtain
\begin{gather}
M\sim\frac{1}{l_P\omega}.
\label{am40}
\end{gather}
Note that in the most ultraviolet case, when $\omega\sim l_P^{-1}$,
we have the natural result $M\sim1$.

To obtain the change of the amplitude per unit time one should
multiply the quantity (\ref{am40}) by the number of elementary
events per unit time $\left(n_{\nu_{\mu}}^{{\cal I}}l_P^2\right)$:
\begin{gather}
\frac{\d}{\d t}M\sim \frac{\hbar c^2l_P}{\omega}n_{\nu_{\mu}}^{{\cal
I}}. \label{am50}
\end{gather}
The comparison of the last result with the well known formulas (see
\cite{1}, \cite{2}) leads to the estimation
\begin{gather}
\frac{\hbar c^2l_P}{\omega}n_{\nu_{\mu}}^{{\cal I}}\sim \frac{\delta
m^2c^4}{\hbar\omega} \rightarrow  n_{\nu_{\mu}}^{{\cal I}}\sim
\frac{c^2\delta m^2}{\hbar^2l_P}\sim10^{42}\mbox{cm}^{-3}.
\label{am60}
\end{gather}
More likely, the estimation (\ref{am60}) is inflated by many orders.

Finally, we conclude that the neutrino oscillations should be
observed since there are mutual transitions of the electron and muon
neutrinos with fixed and equal momenta.

{\bf 6.}  Instead of a Conclusion I would like to pose the question:
Does the effect of neutrino oscillations give evidence in favour of
the space-time granularity?

\begin{acknowledgments}

I thank V. Kiselev for consultations on the neutrino oscillations
problem and S.S. Vergeles for the help in the implementation of some
calculations. This work was supported by SS-3139.2014.2.
\end{acknowledgments}


\end{document}